\documentclass[aps,preprint,a4paper,showpacs,showkeys,superscriptaddress,nofootinbib]{revtex4-1}

\usepackage{latexsym}
\usepackage{amsmath,amssymb}
\usepackage{graphicx}
\usepackage{subfigure}
\usepackage{xcolor}
\usepackage{cancel}
\usepackage{multirow,tabularx}

\newcolumntype{C}[1]{>{\centering\arraybackslash}p{#1}}
\usepackage{hyperref}     
\hypersetup{colorlinks,%
  citecolor=blue,%
  linkcolor=cyan,%
  pdftex}

\usepackage[titletoc]{appendix}
\usepackage{enumerate}

\begin{document}


\title{Hawking-Page phase transition of the Schwarzschild AdS black hole
with the effective Tolman temperature}

\author{Hwajin Eom}%
\email[]{um16@sogang.ac.kr}%
\affiliation{Department of Physics, Sogang University, Seoul, 04107,
  Republic of Korea}%
\affiliation{Center for Quantum Spacetime, Sogang University, Seoul 04107, Republic of Korea}%

\author{Sojeong Jung}%
\email[]{jsquare@sogang.ac.kr}%
\affiliation{Department of Physics, Sogang University, Seoul, 04107,
  Republic of Korea}%
\affiliation{Center for Quantum Spacetime, Sogang University, Seoul 04107, Republic of Korea}%

\author{Wontae Kim}%
\email[]{wtkim@sogang.ac.kr}%
\affiliation{Department of Physics, Sogang University, Seoul, 04107,
	Republic of Korea}%
\affiliation{Center for Quantum Spacetime, Sogang University, Seoul 04107, Republic of Korea}%

\date{\today}

\begin{abstract}
A sufficiently large Schwarzschild AdS black hole with the Hawking temperature
has a positive heat capacity and undergoes the Hawking-Page phase transition.
Even though an arbitrary isothermal surface of a cavity
is introduced, the essential features about the stability and phase transition of the
Schwarzschild AdS black hole remain the same.
In this paper, we study the stability and the Hawking-Page phase transition of the Schwarzschild AdS black hole by employing an alternative local temperature satisfying the Hartle-Hawking vacuum condition which states that
the ingoing and outgoing fluxes vanish at the horizon so that the local temperature is naturally zero there.
The different definition of the local temperature based on the stress tensor approach
provides different types of stability and phase transition.
For a AdS curvature radius below a critical one,
the medium-sized black hole is found to be stable and the other small and large black holes turn
out to be unstable regardless of the AdS curvature radius.
Furthermore, we encounter various critical temperatures.
The first one is for the  Hawking-Page phase transition between the medium-sized black hole and thermal AdS, and
the second one is for a phase transition between the large black hole and thermal AdS; however, the latter one is not essential because
the large black hole and thermal AdS eventually collapse to the medium-sized black hole.
Interestingly, we find an additional critical temperature related to the zeroth-order
phase transition between the medium-sized black hole and thermal AdS.
Consequently, the medium-sized black hole undergoes both the Hawking-Page phase transition
and the zeroth-order phase transition.
\end{abstract}

%



\maketitle


\raggedbottom

\section{introduction}
\label{sec:introduction}
During the last fifty years,
black hole thermodynamics has been one of the most intriguing subjects in
theoretical physics.
In the seminal work by Hawking and Page \cite{Hawking:1982dh},
a sufficiently large Schwarzschild anti-de Sitter (AdS) black hole with the Hawking temperature was shown
to be in stable equilibrium with thermal radiation.
They also demonstrated the so-called Hawking-Page phase transition between the stable
large black hole and thermal AdS
in the canonical ensemble,
which opened the study on the thermodynamic phase transitions of black holes.
Since then,
black hole thermodynamics in AdS space has been
considered as an interesting subject in its own right \cite{Brown:1994gs,Louko:1996dw,Kastor:2009wy,Banerjee:2011au,Kubiznak:2012wp,Xu:2014kwa,Miao:2016ulg,Wang:2018xdz} as well as
in the aspect of holography \cite{Witten:1998qj,Chamblin:1999tk,Caldarelli:1999xj,Nojiri:2001aj,Papantonopoulos:2003wq}.
Some related topics have also been discussed extensively
\cite{Frolov:1996hd,Cai:2001dz,Cvetic:2001bk,Quevedo:2008xn,Wei:2012ui,Cai:2013qga,Yerra:2021hnh}.
The essential reason for the existence of the canonical ensemble
in asymptotically AdS space would be that
the gravitational potential in AdS space
plays the role of a reflecting boundary condition
which makes the black hole be in equilibrium with thermal radiation.

As another approach to imposing the reflecting boundary condition,
the black hole thermodynamics in a cavity was first proposed by York \cite{York:1986it}.
In this approach, an observer is located at the boundary of the cavity,
so one should adopt the local Tolman temperature in thermodynamic analysis \cite{Tolman:1930zza,Tolman:1930ona}.
Consequently,
the asymptotically flat Schwarzschild black hole in the cavity turned out to have
the Hawking-Page phase transition.
Subsequently, black hole thermodynamics and the Hawking-Page transition have been studied
for other types of black holes surrounded by the cavity \cite{Braden:1990hw,Carlip:2003ne,Zaslavskii:2003cr,Lundgren:2006kt,Gim:2014ira,Wang:2020hjw}.

In the Schwarzschild AdS black hole, what would happen if one introduces the Tolman temperature
instead of the Hawking temperature.
It means that one should investigate black hole thermodynamics by using the local temperature defined at an arbitrary
distance from the Schwarzschild AdS black hole; however,
any local observer would find a similar thermodynamic result \cite{Peca:1998cs,Akbar:2004ke}.
Of course, even in the limit of an infinite AdS curvature radius when
the black hole approaches the Schwarzschild black hole surrounded by a cavity,
the Hawking-Page phase transition would occur.

On the other hand, for the Schwarzschild AdS black hole,
there is an alternative local temperature
so called the effective Tolman temperature~\cite{Eune:2017iab}.
One of the advantages of the effective Tolman temperature is indeed obeying the Hartle-Hawking vacuum condition
compatible with the black hole in thermal equilibrium so that the black hole
can be successfully described as a hot object immersed in the sea of
Hawking particles. In addition, the effective Tolman temperature reflects the fact that the Hawking radiation is of relevance to
the trace anomaly \cite{Christensen:1977jc}, which is different from the Tolman temperature assuming
the traceless stress tensor~\cite{Tolman:1930zza,Tolman:1930ona}.

In this paper, we will investigate stability and phase transition of
the Schwarzschild AdS black hole by using the
effective Tolman temperature.
For the AdS curvature radius below a critical one,
the heat capacity is found to be positive for a medium-sized black hole while
it is negative for the other small and large black holes regardless of the AdS curvature radius.
We shall have three critical temperatures: the first one is for the Hawking-Page phase transition between the medium-sized black hole and thermal AdS, and
the second one is for a phase transition between the large black hole and thermal AdS.
However, the second critical temperature is not essential because
the large black hole and thermal AdS eventually collapse to the medium-sized black hole.
Lastly, the third critical temperature is found to be of relevance to the zeroth-order
phase transition \cite{Gunasekaran:2012dq,Altamirano:2013ane,Kubiznak:2015bya} between the medium-sized black hole and thermal AdS.
Consequently, we find that the medium-sized black hole undergoes the Hawking-Page phase transition
at the first critical temperature and the zeroth-order phase transition at the third
critical temperature.

The organization of this paper is as follows.
In Sec.~\ref{sec:temperature},
we will explain how the effective Tolman temperature can be obtained
from the Stefan–Boltzmann law induced by the trace anomaly.
In Sec.~\ref{sec:stability}, the heat capacity will be calculated explicitly
and then the medium-sized black hole will be shown to be stable.
In Sec.~\ref{sec:transition},
we will obtain the free energy and then study the Hawking-Page phase transition and
zeroth-order phase transition between Schwarzschild AdS black hole and thermal AdS.
Finally, conclusion and discussion will be given in Sec.~\ref{sec:conclusion}.

\section{Effective Tolman temperature}
\label{sec:temperature}
From the trace anomaly of quantum stress tensor for a scalar field,
we present the derivation of the effective Tolman temperature appropriate to the Schwarzschild AdS black hole.
For this purpose, we choose a tractable two-dimensional Schwarzschild AdS black hole without loosing any essential
thermodynamic properties of a four-dimensional Schwarzschild AdS black hole, so the line element is given as
    \begin{equation}
    \label{eq:metric}
    ds^2=-f(r) dt^2 +\frac{dr^2}{f(r)},
    \end{equation}
where the metric function is $f(r)=1-2M/r +r^2/\ell^2$
with $M$ and $\ell$ being the black-hole mass and the AdS curvature radius.
The metric function can be rewritten as
$f(r)=\left( 1-r_h/r\right) \left[\left(r^2+r r_h+r_h^2\right)/\ell^2 +1\right]$,
where $r_h$ is a black hole horizon related to $M$ through
    \begin{equation}
    \label{eq:mass_relation}
    M=\frac{r_h}{2}\left(\frac{r_h^2}{\ell^2}+1 \right).
    \end{equation}
Note that the black hole mass is monotonically increasing with respect to the black hole horizon.
The Hawking temperature is also obtained from the surface gravity at the horizon as
\begin{equation}
\label{HT}
T_H = \frac{3 r_h^2+\ell^2 }{4\pi\ell^2 r_h }.
\end{equation}
For $\ell \to \infty$ in Eqs.~\eqref{eq:metric}, \eqref{eq:mass_relation}, and \eqref{HT},
the Schwarzschild AdS black hole has a well-defined Schwarzschild limit.

First of all, we now determine the stress tensor responsible for thermal particles of heat reservoir.
Let us consider a two-dimensional quantum
scalar field on the classical background \eqref{eq:metric}, then its trace anomaly is given as \cite{Deser:1993yx,Duff:1993wm}
    \begin{equation}
    \label{eq:anomaly}
    T^\mu_\mu = \frac{1}{24\pi} R
    \end{equation}
which is proportional to the curvature scalar
$R=-f'' = (2 r_h/r^3) (r_h^2/\ell^2+1 )-2/\ell^2$,
where the prime denotes the derivative with respect to $r$.
From the trace anomaly \eqref{eq:anomaly}
and the covariant conservation law of the stress tensor of $\nabla_\mu T^{\mu\nu}=0$,
the explicit form of the stress tensor, in the conformal gauge of $ds^2 =-e^{2\rho(\sigma^+,\sigma^- )} d\sigma^+ d\sigma^-$ where $e^{2\rho} =f(r)$
and $\sigma^{\pm}=t \pm \int^{r^*} (dr/f(r))$,
can be determined as
    \begin{equation}
    \label{eq:em_tensor}
    T_{\pm\pm} =\frac{1}{96\pi}\left[ ff''-\frac 1 2 f'^2+t_\pm \right],\quad
    T_{+-} =\frac{1}{96\pi} f f'',
    \end{equation}
where $t_\pm$ reflect the nonlocality of the trace anomaly.
In thermal equilibrium, the ingoing and outgoing fluxes are the same: $T_{++}=T_{--}$, so
the net flux is zero. In the Hartle-Hawking vacuum condition \cite{Hartle:1976tp,Israel:1976ur},
one can describe the black hole in thermal equilibrium by choosing $t_\pm$ as
    \begin{equation}
    \label{eq:Hartle-Hawking}
    t_\pm =\frac 1 2 f'(r_h)^2 =\frac 1 2\left(\frac{3 r_h}{\ell^2}+\frac{1}{r_h} \right)^2.
    \end{equation}

Next, for a static observer located at the radius $r$ of the boundary,
the two-velocity is solved as $u^\mu =\left(1/\sqrt{f(r)},0 \right)$
which is written as
$u^\pm=1/\sqrt{f(r)}$ in the conformal gauge.
Then, from Eqs.~\eqref{eq:em_tensor} and \eqref{eq:Hartle-Hawking}, we get
the proper energy density  in the Hartle-Hawking state as
    \begin{eqnarray}
    \label{eq:proper_energy density}
    \epsilon=T_{\mu\nu}u^\mu u^\nu
        =\frac{1}{96\pi f} \left[ 4 f f''-f'^2+ f'(r_h)^2 \right].
    \end{eqnarray}
On the other hand, the effective Tolman temperature can be read off from the modified Stefan-Boltzmann law as \cite{Gim:2015era}
    \begin{equation}
    \label{eq:SB_law_energy}
    \gamma T^2=\epsilon  +\frac 1 2 T^\mu_\mu,
    \end{equation}
where $\gamma=\pi/6$ for a single scalar field.
In the absence of trace anomaly,
Eq.~\eqref{eq:SB_law_energy}
reduces to the conventional Stefan-Boltzmann law  so that it reproduces the Tolman temperature.

Plugging the trace anomaly \eqref{eq:anomaly} and the proper energy density \eqref{eq:proper_energy density} into the modified
Stefan-Boltzmann law \eqref{eq:SB_law_energy}, we obtain the following effective Tolman temperature
at $r$ \cite{Eune:2017iab}
    \begin{equation}
    \label{eq:eff_temperature}
    T=\frac{1}{4\pi r}\frac{\sqrt{(1-x) \left(x^2+\alpha^2\right) \left[3x^2 \left(3+2x+x^2\right)+\alpha^2 \left(1+2x+3x^2\right)\right]}}{\alpha x\sqrt{1+\alpha^2 +x+x^2}},
    \end{equation}
where $x=r_h/r$ and $\alpha=\ell/r$ with $0<x<1$. Especially, $x=r_h$ and $\alpha =\ell$ at $r=1$.
As expected, the effective Tolman temperature~\eqref{eq:eff_temperature} naturally vanishes at the horizon of $x=1$
since the black hole is in the Hartle-Hawking vacuum state which requires that the ingoing and outgoing fluxes
should vanish at the horizon.
This feature of the effective temperature is quiet different
from that of the usual Tolman temperature divergent at the horizon,
so the thermodynamic stability of the black hole would be different.
In Eq.~\eqref{eq:eff_temperature}, it is worth noting that
the Schwarzschild limit can be obtained by taking $\ell\to\infty$, $i.e.$, $\alpha\to\infty$ for a fixed $r$.
Thus, the effective Tolman temperature becomes $T|_{\ell\to\infty} = (1/4\pi r_h) \sqrt{(1-x)\left(1+2x+3x^2\right)}$.
At infinity, it approaches the Hawking temperature of the Schwarzschild black hole: $T|_{\ell\to\infty} = 1/(4\pi r_h)$.

On the other hand, in the case of an asymptotically flat
Schwarzschild black hole,
the Tolman temperature of $T_H/\sqrt{f(r)}$
approaches the Hawking temperature at infinity.
In particular,
the Tolman temperature can be derived from the Stefan-Boltzmann law of $ \gamma T^2 =\epsilon$.
Note that the Tolman's original formulation \cite{Tolman:1930zza,Tolman:1930ona} rests upon a traceless stress tensor for matter fields.
However, the stress tensor on the black hole background receives quantum correction and its trace is non-vanishing quantum-mechanically.
In fact, the trace anomaly is also responsible for Hawking radiation \cite{Christensen:1977jc}.
Thus, the original Tolman's formulation should be extended to incorporate the non-vanishing trace of stress tensor
and so the Stefan-Boltzmann law should be appropriately modified as $\gamma T^2=\epsilon +(1/2) T^\mu_\mu$ \cite{Eune:2017iab}.
At infinity, the effective Tolman temperature reduces to the original one because the trace anomaly vanishes there,
which means that
there is no difference between the behaviors of the effective and original ones in the Schwarzschild black hole.
However, the above argument is not valid for a Schwarzschild AdS black hole because the trace anomaly proportional to
the curvature scalar persists even at infinity.
Explicitly, the effective Tolman temperature \eqref{eq:SB_law_energy} at infinity
can be written as $\gamma T^2 |_{r\to\infty} =\epsilon |_{r\to\infty}+(1/2) T^\mu_\mu |_{r\to\infty} = 0$, where
$\epsilon |_{r\to\infty}=1/(24\pi\ell^2)$ and $T^\mu_\mu |_{r\to\infty} =-1/(12\pi\ell^2)$ in our model
so that the effective Tolman temperature vanishes and it does not approach the Hawking temperature.
In some sense, the vanishing temperature at infinity appears to be
plausible since there is an infinite potential barrier at the AdS boundary
so that the amplitude of matter fields vanishes like the
box model in quantum mechanics.
It would be interesting to note that the local temperature derived from the Euclidean action formalism \cite{York:1986it} and the quasi-local formalism \cite{Brown:1994gs} takes
the well-known Tolman's form but it is also zero at infinity thanks to the redshift factor.

\section{Thermodynamic stability}
\label{sec:stability}
We calculate the heat capacity by using the effective Tolman temperature \eqref{eq:eff_temperature}
in order to study the thermodynamic stability of the Schwarzschild AdS black hole
surrounded by an isothermal surface of a cavity placed at $r$.
Now, the entropy can be defined as the Bekenstein-Hawking entropy as
    \begin{equation}
    \label{eq:entropy}
    S=\pi r_h^2
    \end{equation}
which is independent of the cavity size
in quasilocal thermodynamics
since the degrees of freedom of the black hole would be irrelevant to the boundary of the cavity \cite{Brown:1994gs}.
From the thermodynamic first law, the thermodynamic energy can be obtained as
    \begin{equation}
    \label{eq:tot_energy}
    E (M) \!\!=\!\! \int_0^M \!\!T(M)dS(M) \!\!=\!\!\frac{r}{2\alpha}\!\! \int_0^x \!\frac{\sqrt{(1 \! -\!x) \left(x^2 \!+\! \alpha^2\right) \left[3x^2 \left(3 \!+\! 2x \!+\! x^2\right)+\alpha^2 \left(1 \!+\! 2x \!+\! 3x^2\right)\right]}}{\sqrt{1+\alpha^2 +x+x^2}} dx,
    \end{equation}
where Eqs.~\eqref{eq:eff_temperature} and \eqref{eq:entropy} are used through Eq.~\eqref{eq:mass_relation}.
Then, from Eqs.~\eqref{eq:eff_temperature} and \eqref{eq:tot_energy}, the heat capacity is calculated as
    \begin{equation}
    \label{eq:heat_capacity}
    C_V =\left(\frac{\partial E}{\partial T}\right)_r
    = \left(\frac{\partial E}{\partial x}\right)_r \left(\frac{\partial T}{\partial x}\right)_r^{-1},
    \end{equation}
where
    \begin{equation}
    \label{eq:find_extrema}
    \left(\frac{\partial T}{\partial x}\right)_r =-\frac{(3x^2+\alpha^2)D(x)}{8\pi\alpha x^2 \sqrt{A(x)} B(x)\sqrt{B(x)}},
    \end{equation}
which consists of some polynomials defined as
    \begin{align}
    A(x) =&(1-x) \left(x^2+\alpha^2\right) \left[3x^2 \left(3+2x+x^2\right)+\alpha^2 \left(1+2x+3x^2\right)\right],\\
    B(x) =&1+\alpha^2 +x+x^2,\\
    \label{eq:poly}
    D(x)=&2\alpha^2 (1+\alpha^2)+\alpha^2 (4+\alpha^2)x-6x^2+3\alpha^2(2+\alpha^2)x^3\\
    \notag &+6(1+\alpha^2) x^4 +3(3+2\alpha^2)x^5+6x^6+3x^7,
    \end{align}
where $0<x<1$.
Note that the thermodynamic stability of the black hole relies on $(\partial T/\partial x)_r$
because $(\partial E/\partial x)_r$ is positive definite.
Explicitly, $A(x)$ and $B(x)$ in Eq. \eqref{eq:find_extrema} are always positive so that
the sign of $(\partial T/\partial x)_r$ is ascribable to that of $D(x)$.
\begin{figure}[b]
\centering
\subfigure[]{\includegraphics[width=0.38\textwidth]{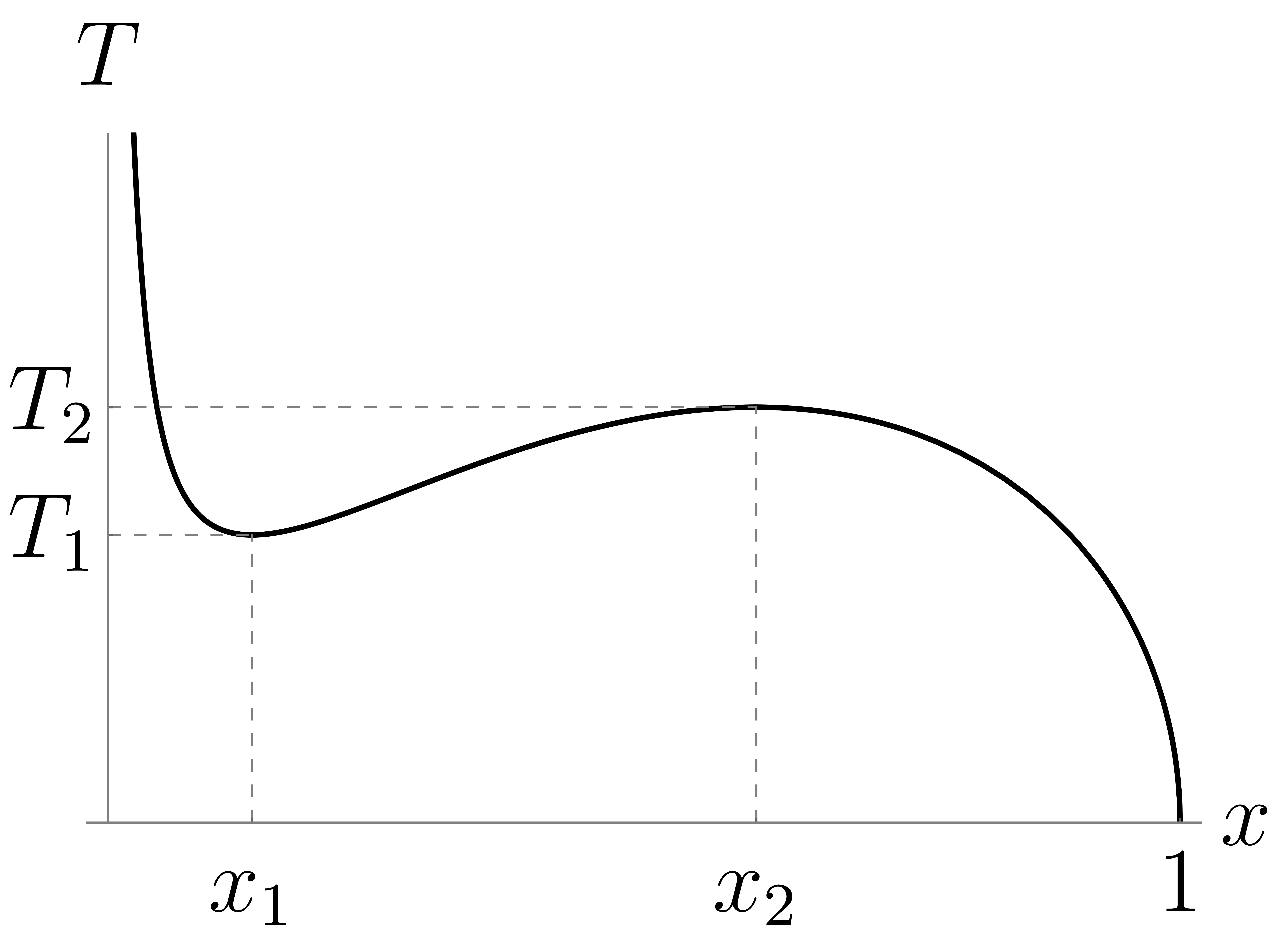}\label{fig:temperature}}
\subfigure[]{\includegraphics[width=0.38\textwidth]{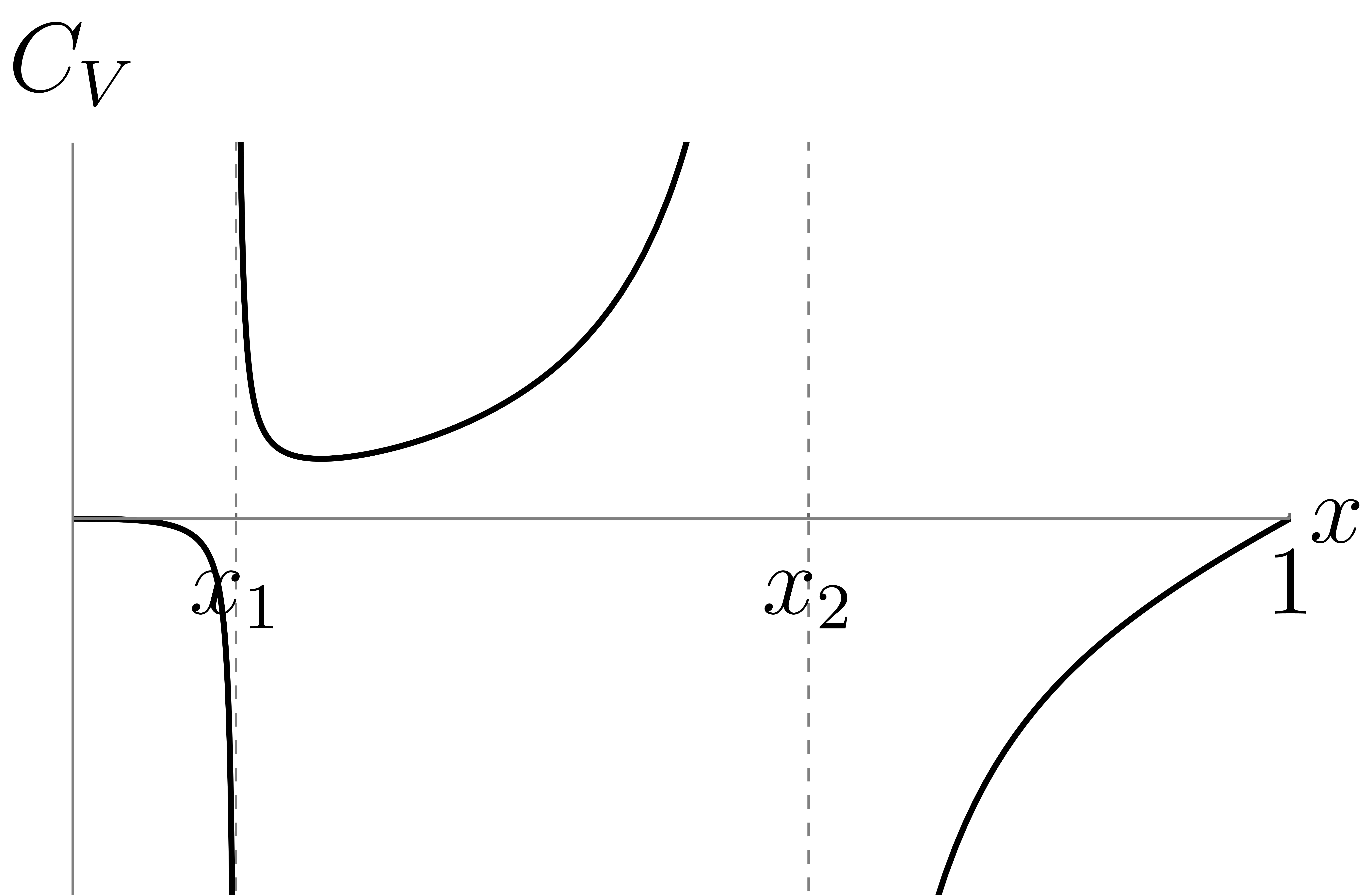}\label{fig:heat_capacity}}
\caption{The effective Tolman temperature
\eqref{eq:eff_temperature} and the heat capacity \eqref{eq:heat_capacity} are plotted
for $\alpha=0.2 < \alpha^*=0.3867$ and $r=1$.
In Fig.~\ref{fig:temperature}, there exist two extreme values of $T_1$ and $T_2$
at $x_1$ and $x_2$. In Fig.~\ref{fig:heat_capacity},
the heat capacity is singular at $x_1$ and $x_2$ and it is positive for $x_1 <x< x_2$; otherwise,
it is negative.}
\end{figure}

Using Eq.~\eqref{eq:find_extrema}, we first find extrema of the temperature in order to figure out the sign
of the heat capacity.
Rewriting $D(x)$ in terms of a power of $\alpha^2$ as
$D(x) = d_0(x) + d_1(x) \alpha^2+d_2(x) \alpha^4$,
where $d_0(x)= -6 x^2 +6x^4 +9x^5 +6x^6 +3 x^7$, $d_1(x)=2+4x+ 6x^3 +6x^4 +6 x^5$, and $d_2(x)=2+x+3x^3$,
one can easily find the roots of $D(x)$ as
    \begin{equation}
    \label{eq:alpha_square}
    \alpha^2 = \frac{-d_1(x) \pm\sqrt{d_1^2 (x)-4 d_0 (x)d_2 (x)}}{2 d_2 (x)},
    \end{equation}
where $d_1(x)$ and $d_2(x)$ are positive.
To make $\alpha$ real, let us choose the plus sign in Eq.~\eqref{eq:alpha_square}.
As a result, the number of roots of $D(x)$ is obtained as
    \begin{equation}
    \label{eq:zeros}
    \textrm{the number of roots}=
    \begin{cases}
    2& \textrm{for }0<\alpha<\alpha^*,\\
    1& \textrm{for }\alpha=\alpha^*,\\
    0& \textrm{for }\alpha>\alpha^*,
    \end{cases}
    \end{equation}
where the critical ratio between the AdS curvature radius and the cavity size is numerically found as approximately
$\alpha^*=0.3867$. Note that the critical value is independent of the cavity size.

In the case of $\alpha < \alpha^*$,
the effective Tolman temperature \eqref{eq:eff_temperature} and the heat capacity \eqref{eq:heat_capacity} are plotted in Figs.~\ref{fig:temperature} and \ref{fig:heat_capacity}, respectively.
The effective Tolman temperature has two extrema of $T_1$ and $T_2$
at $x_1$ and $x_2$.
Accordingly, the heat capacity suffers from
discontinuities and changes its sign at $x_1$ and $x_2$
so that a large black hole in $x>x_2$ ($M>M_2)$ as well as a small black hole in $x<x_1$ ($M<M_1)$
turns out to be unstable.
In $x_1 <x< x_2$ ($M_1 <M<M_2$), the heat capacity is positive, so the medium-sized Schwarzschild AdS black hole is stable.
However, in the case of $\alpha  >\alpha^*$, the heat capacity \eqref{eq:heat_capacity} is always negative,
so any size of the Schwarzschild AdS black hole is unstable.
In particular, for the degenerate case of $\alpha =\alpha^*$, there exists one extremum but
the heat capacity is negative because the positive part of the heat capacity disappears when $x_1 \to x_2$.
On the other hand, in the Schwarzschild limit of $\alpha \to \infty$,
the heat capacity \eqref{eq:heat_capacity} reduces to
    \begin{equation}
    \label{eq:heat_ell_infinity}
    C_V|_{\ell\to\infty} =-\frac{4\pi r x^2 (1-x)\left(1+2x+3x^2 \right)}{2+x+3x^3},
    \end{equation}
so the Schwarzschild black hole must be unstable.
This result is quiet distinct from
the fact that a sufficiently large Schwarzschild black hole is stable when the conventional Tolman temperature is used.

\section{Hawking-Page phase transition}
\label{sec:transition}

\begin{figure}[b]
\centering
\subfigure[]{\includegraphics[width=0.41\textwidth]{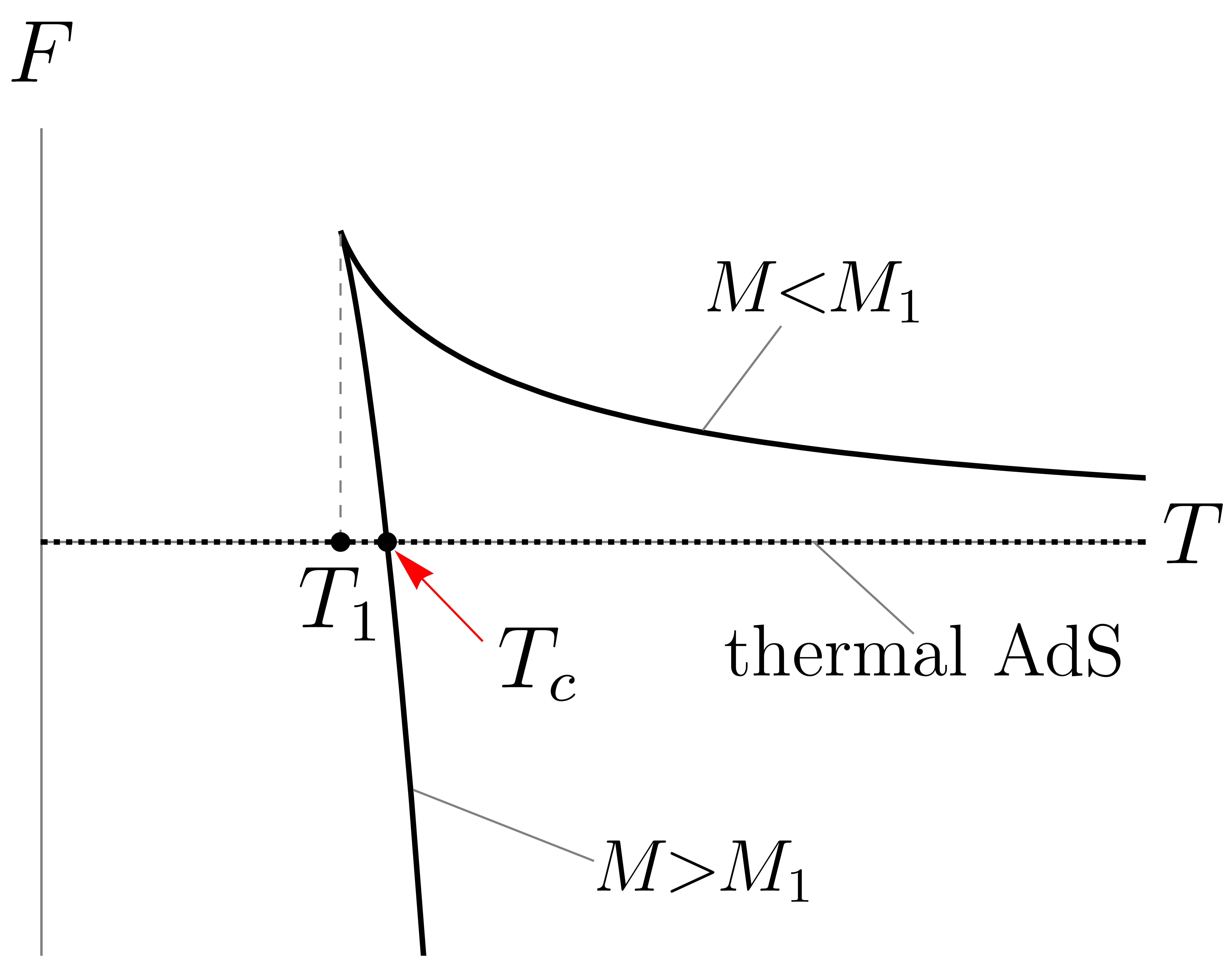}\label{fig:free_conventional}}
\subfigure[]{\includegraphics[width=0.57\textwidth]{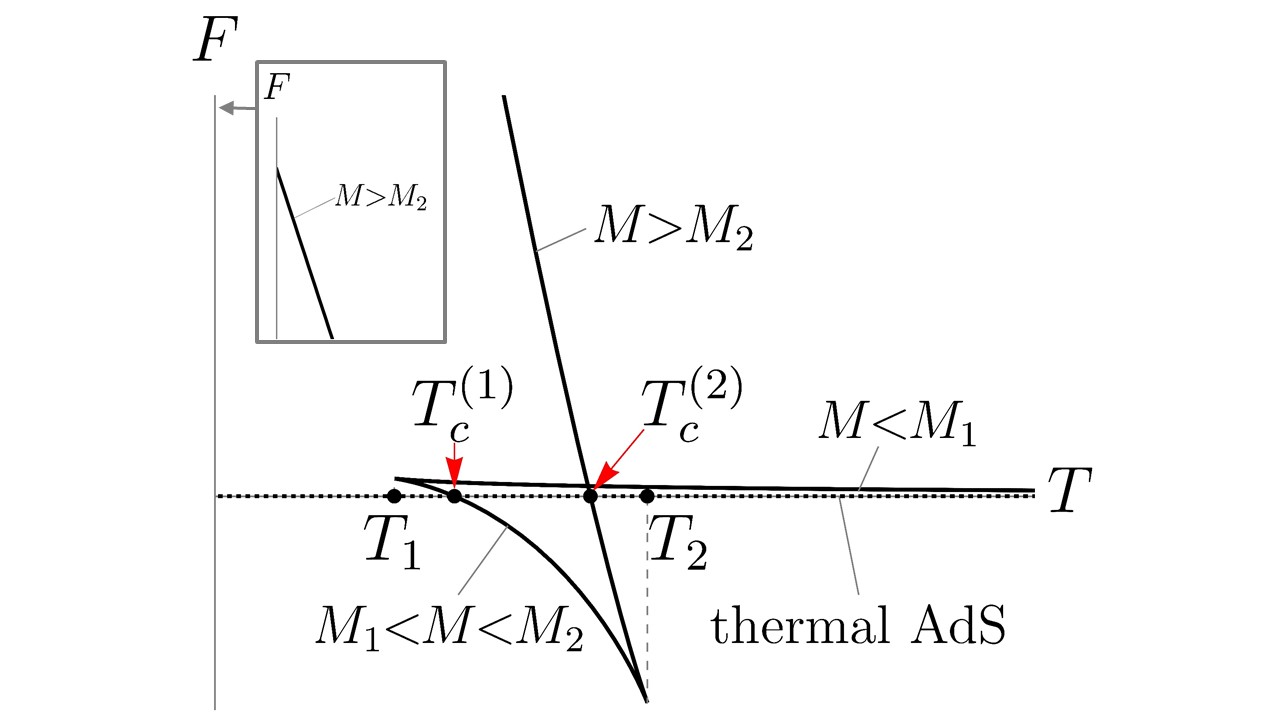}\label{fig:free-energy}}
\caption{
In Figs.~\ref{fig:free_conventional} and \ref{fig:free-energy},
the solid curves describe the free energies \eqref{eq:free_conventional} and \eqref{eq:free-energy} for $\alpha=0.2 <\alpha^*$ and $r=1$.
The small dotted horizontal lines of $F=0$ are for
the free energy of thermal AdS.
In Fig.~\ref{fig:free_conventional}, there exists a cusp  at the minimum temperature of $T=T_1$,
and $T_c$ is a critical temperature.
In Fig.~\ref{fig:free-energy},
$T_1$ and $T_2$ are the two extreme values in Fig.~\ref{fig:temperature},
and $T_c^{(1)}$ is a critical temperature for the Hawking-Page transition
while $T_c^{(2)}$ is an apparent critical temperature.
Note that the zeroth-order phase transition occurs at $T_2$.
For a large black hole for $M>M_2$, the free energy plotted in the small box in Fig.~\ref{fig:free-energy}
shows that it reaches the maximum value of $F=1.084$
at $T=0$.
}
\end{figure}
Let us first investigate whether the Hawking-Page phase transition
between the Schwarzschild AdS black hole and
thermal AdS can be modified or not
when the conventional Tolman temperature is used.
In this case, from the thermodynamic
first law with the Tolman temperature of $T=T_H /\sqrt{f(r)}$, the local thermodynamic energy is calculated as
\begin{equation}
E=\frac{r}{\alpha} \left[\sqrt{1+\alpha^2} -\sqrt{1+\alpha^2-\alpha^2 x-x^3} \right],
\end{equation}
and the free energy of the Schwarzschild AdS black hole is easily obtained as
    \begin{equation}
    \label{eq:free_conventional}
    F=E-TS=\frac{r}{\alpha} \left(\sqrt{1+\alpha^2}-\frac{4+4\alpha^2-3 \alpha^2 x-x^3}{4\sqrt{1+\alpha^2-\alpha^2 x-x^3}} \right),
    \end{equation}
which is plotted in Fig.~\ref{fig:free_conventional}
with respect to the Tolman temperature.
As expected, there is a single critical temperature for an arbitrary $\alpha$:
the large black hole of $M>M_1$ decays into thermal AdS for $T_1 <T<T_c$ and
thermal AdS is nucleated for $T > T_c$. Although the small black hole of $M <M_1$ for $T_1 <T<T_c$ acquires mass
from the heat reservoir and then  it becomes a large black hole, but it eventually decays into thermal AdS.
In particular, taking $\alpha \to \infty$, $i.e.$, $\ell \to \infty$ for a finite $r$,
one can retrieve York's free energy of the asymptotically flat Schwarzschild black hole \cite{York:1986it}.
Hence, the Hawking-Page phase transition of the Schwarzschild AdS black hole occurs irrespective of
the size of the AdS radius $\ell$.

We are now in a position to study how the effective Tolman temperature works in the stability and
the phase transition of the Schwarzschild AdS black hole.
From Eqs.~\eqref{eq:eff_temperature}, \eqref{eq:entropy}, and  \eqref{eq:tot_energy},
the free energy of our interest is written as
    \begin{equation}
    \label{eq:free-energy}
    F =E -T S =\frac{r}{2\alpha} \int_0^x \frac{\sqrt{A(x)}}{\sqrt{B(x)}}dx
    -\frac{r}{4\alpha}\frac{x\sqrt{A(x)}}{\sqrt{B(x)}},
    \end{equation}
which is plotted in Fig.~\ref{fig:free-energy}.
For $\alpha <\alpha^*$, the three types of black holes were introduced in Sec.~\ref{sec:stability}:
a small black hole in $M<M_1$,
a medium-sized black hole in $M_1<M<M_2$,
and a large black hole in $M>M_2$.
Accordingly,
the corresponding free energies can be divided into three branches.
In Fig.~\ref{fig:free-energy},
we find three critical temperatures such as $T_c^{(1)}$, $T_c^{(2)}$ and $T_2$.
There exist a critical temperature of $T_c^{(1)}$ at which
the phase transition occurs between the medium-sized black hole and thermal AdS
and another critical temperature of $T_c^{(2)}$ between the large black hole and thermal AdS.
Specifically, {\bf (i)} for $T <T_c^{(1)}$, the free energies of the small, medium-sized, and large black holes are positive,
so all of them decay into thermal AdS.
The medium-sized black hole is at least locally stable for $T_1<T <T_c^{(1)}$ as seen from the calculation of the heat capacity
in Fig.~\ref{fig:heat_capacity}, but thermal AdS is eventually more probable.
{\bf (ii)} For $T_c^{(1)} <T <T_{2}$, the medium-sized black hole is shown to be not only stable but also
the most probable state
and so the small and large black holes as well as thermal AdS must eventually collapse into the medium-sized black hole.
Therefore, $T_c^{(2)}$ must be an apparent critical temperature unlike
the genuine critical temperature $T_c$ in Fig.~\ref{fig:free_conventional}.
More importantly, at $T = T_2$, there is a zeroth-order phase transition \cite{Gunasekaran:2012dq,Altamirano:2013ane,Kubiznak:2015bya}
between the medium-sized black hole and thermal AdS
in which a discontinuity between the global minimum of $F$ exists.
{\bf (iii)} For $T >T_2$,
the small black hole decays into thermal AdS.


\section{conclusion and discussion}
\label{sec:conclusion}
We studied the thermodynamic stability and the phase transition
of the Schwarzschild AdS black hole in a cavity.
The key ingredient of our calculations was based on
the effective Tolman temperature induced by the trace anomaly of the
stress tensor for a massless scalar field.
Of course, in the traceless limit, the effective Tolman temperature reduced to
the usual Tolman temperature which reproduces the well-known
Hawking-Page phase transition.
In our paper, the effective Tolman temperature was found to have two extrema, and
thus, the heat capacity has two singularities.
Importantly, the effective Tolman temperature shares most of features of the
Tolman temperature for the small black hole, but
it is drastically different from the behaviors of the large black hole
in the sense that the effective Tolman
temperature vanishes when
the black hole size becomes comparable to the cavity size. This vanishing temperature at the horizon is consistent with
the Hartle-Hawking state in thermal equilibrium with Hawking radiation.

On the other hand, the non-trivial behavior of the effective Tolman temperature depends on
the ratio between $\ell$ and $r$ so that the stable black hole phase happens for $\alpha<\alpha^*$
while the black holes in $\alpha\ge\alpha^*$ including
the Schwarzschild limit for which $\ell \to \infty$
are unstable.
One of the most interesting things to distinguish
from the conventional thermodynamic stability analysis is that the stable equilibrium is realized only
when the black hole size is in a finite range of $M_1 <M<M_2$.
This feature is also quiet different from the fact that a large Schwarzschild black hole in a cavity is stable.
Finally, from the free energy analysis,
the three critical temperatures of $T_c^{(1)}$, $T_c^{(2)}$ and $T_2$ were found, but
the medium-sized black hole in $T_c^{(1)}<T <T_2$ was turned out to be the most probable state.

The previous studies showed that the Schwarzschild black hole can be made stable either by introducing the cavity
or by immersing the black hole in AdS space since
the cavity plays the role of AdS boundary
as a reflecting boundary condition. At first sight,
in the Schwarzschild black hole, it seems to be redundant to consider AdS space together with
the cavity. However, the present study shows that
the cavity and AdS space should be accommodated simultaneously to make
the Schwarzschild black hole stable if the effective Tolman temperature is used as a local temperature.
In addition, the AdS curvature radius and the cavity size interfere in our model,
so the value of the ratio between them is crucial in our stability analysis.
It would be interesting to clarify further
the reason why the cavity appears to be insufficient to play the role of AdS boundary in our formulation.

Finally, Eq.~\eqref{eq:tot_energy} comes from the thermodynamic first law and it
plays a key role to our work.
Thus, one might wonder how to justify the thermodynamic first law
by using the Iyer and Wald's covariant construction \cite{Iyer:1994ys} or
the on-shell Euclidean action method \cite{York:1986it}.
In the former method, at a bifurcation point,
the energy and entropy were obtained from Noether charges associated with
diffeomorphism symmetry of Lagrangian and then the Hawking temperature was identified with the surface gravity which plays a role of
a proportional constant in the thermodynamic first law.
As long as one is concerned with conserved quantities derived from the diffeomorphism symmetry,
the Hawking temperature naturally appears in the thermodynamic first law.
But the surface gravity at the event horizon cannot be a local temperature for an observer
located at the boundary of a cavity. In addition, the effective Tolman temperature was
induced from the trace anomaly of matter fields on the black hole background.
To accommodate the local temperature induced by the trace anomaly into the regime of covariant formalism,
we may have to generalize or modify the Iyer and Wald's construction in a certain way.
In the latter method,
the Tolman temperature consistent with the thermodynamic first law
was derived from the periodicity of Euclidean
time in the on-shell action with the exact metric solution.
If we think about this in parallel, we need to consider the on-shell action
which consists of the classical and quantum-mechanical effective actions.
For the explicit on-shell Euclidean action, an exact metric solution must be obtained,
which seems to be very hard.
In these respects, this issue deserves further study.

\acknowledgments
This work was supported by the National Research Foundation of Korea(NRF) grant funded by the Korea government(MSIT) (No. NRF-2022R1A2C1002894).
WK was partially supported by Basic Science Research Program through the National Research Foundation of Korea(NRF) funded by the Ministry of Education through the Center for Quantum Spacetime (CQUeST) of Sogang University (NRF-2020R1A6A1A03047877).
HE was partially supported by Basic Science Research Program (NRF-2022R1I1A1A01068833).


\bibliographystyle{JHEP}       

\bibliography{references}

\end{document}